\documentclass[reprint,superscriptaddress,amsmath,amssymb,aps,prl]{revtex4-2}
\usepackage[english]{babel}
\renewcommand{\selectlanguage}[1]{}
\usepackage[utf8]{inputenc}
\usepackage[normalem]{ulem}
\usepackage{physics}
\usepackage{siunitx}
\usepackage{upgreek}
\usepackage[dvipsnames]{xcolor}
\usepackage{graphicx}
\usepackage{dcolumn}
\usepackage{bm}
\definecolor{rred}{RGB}{151,23,46} 
\usepackage[colorlinks=true,citecolor=rred,urlcolor=rred,linkcolor=rred]{hyperref}
\DeclareSIUnit\gal{Gal}

\newcommand{\texps}{T_{\text{exp,ref}}}
\newcommand{\texpL}{T_{\text{exp,sig}}}

\newcommand{\texp}{T_{\text{exp}}}
\newcommand{\tdrop}{T_{\text{drop}}}
\newcommand{\tsplit}{T_{\text{split}}}
\newcommand{\tsep}{T_{\text{sep}}}
\newcommand{\tseps}{T_{\text{sep,ref}}}
\newcommand{\tsepl}{T_{\text{sep,sig}}}

\newcommand{\tsig}{T_{\text{sig}}}
\newcommand{\tref}{T_{\text{ref}}}

\def\ANU{Department of Quantum Science and Technology, The Australian National University, Canberra, ACT 2601, Australia.}
\def\Nomad{\emph{N\textsc{omad} A\textsc{tomics}}, 33 Elizabeth St, Richmond, VIC 3121, Australia}
\begin{document}
\preprint{APS/123-QED}

\title{Dual Open Atom Interferometry for Compact and Mobile Quantum Sensing}

\author{Yosri Ben-A\"{i}cha}
\email{yosri.benaicha@anu.edu.au}
\altaffiliation[Current address: ]{Q-CTRL, 93 Shepherd St, Chippendale, NSW 2008, Australia.}
\affiliation{\ANU}%

\author{Zain~Mehdi}
\affiliation{\ANU}%

\author{Christian~Freier}
\affiliation{\Nomad}%

\author{Stuart~S.~Szigeti}
\altaffiliation[Current address: ]{Q-CTRL, 93 Shepherd St, Chippendale, NSW 2008, Australia.}
\affiliation{\ANU}%

\author{Paul~B.~Wigley}
\affiliation{\Nomad}%

\author{Lorcán~O.~Conlon}
\affiliation{\ANU}%

\author{Ryan~Husband}%
\affiliation{\ANU}%

\author{Samuel~Legge}%
\affiliation{\ANU}%

\author{Rhys~H.~Eagle}%
\affiliation{\ANU}%

\author{Joseph~J.~Hope}
\affiliation{\ANU}%

\author{Nicholas~P.~Robins}%
\altaffiliation[Current address: ]{Q-CTRL, 93 Shepherd St, Chippendale, NSW 2008, Australia.}
\affiliation{\ANU}%

\author{John~D.~Close}
\affiliation{\ANU}

\author{Kyle~S.~Hardman}
\affiliation{\Nomad}%

\author{Simon~A.~Haine} 
\affiliation{\ANU}%

\author{Ryan~J.~Thomas}%
\affiliation{\ANU}%
\date{\today}
\begin{abstract}

We demonstrate an atom interferometer measurement protocol compatible with operation on a dynamic platform. Our method employs two open interferometers, derived from the same atomic source, with different interrogation times to eliminate initial velocity dependence while retaining precision, accuracy, and long term stability. We validate the protocol by measuring gravitational tides, achieving a precision of \SI{4.5}{\micro\gal} in 2000 runs (\SI{6.7}{\hour}), marking the first demonstration of inertial quantity measurement with open atom interferometry that achieves long-term phase stability.
\end{abstract}
\maketitle
Atom interferometers, through their high precision \cite{Peters:2001, Rosi2014, Altin2013, hardman2016,Zhang2023}, accuracy~\cite{Karcher:2018, Morel2020Determination,janvier2022, Geiger:2020}, and stability \cite{Gillot2014, Freier_2016, Menoret:2018, wu2019}, hold tremendous promise as sensors for Earth and climate science~\cite{timmen_observing_2012}, geodesy~\cite{Stockton2011,Geiger2011,Bidel:2020, Bidel:2018, Wu:2023, Becker2018, migliaccio_mocass_2019, Trimeche:2019,Leveque:2021}, mineral exploration~\cite{Evstifeev:2017, Bongs2019}, groundwater mapping and monitoring~\cite{schilling_gravity_2020}, navigation~\cite{Jekeli:2005, Battelier:2016, Cheiney2018, wu2019,Templier:2022,Narducci2022,WangNav:2023}, planetary exploration~\cite{MULLER2020}, and space-based fundamental physics tests~\cite{Dimopoulos:2007,Mutinga2013,Tino:2021,Gaaloul2022,Du2022,Pahl2024Atom}. All require laboratory-grade performance in the field on dynamic mobile platforms. This is a significant challenge and one that motivates this Letter.

In three-pulse ($\pi/2$-$\pi$-$\pi/2$) `closed' atom interferometry, the most widespread measurement protocol, matterwaves in the two interferometer arms perfectly overlap in position and momentum space at the output beamsplitter, and the interferometer phase is inferred from the population difference between the two interferometer outputs~\cite{Kasevich:1992}. These schemes are straightforward to implement in a stable laboratory, but perform poorly in the presence of platform acceleration and rotations, which can degrade sensor accuracy and sensitivity by inducing spurious inertial accelerations and preventing perfect closure of the interferometer~\cite{Templier:2022,de_castanet_atom_2024}. Furthermore, inference of the interferometer phase in general requires multiple interferometer measurements~\cite{Barrett:2014b}, resulting in a trade-off between sensor measurement rate (i.e. bandwidth) and accuracy, which is undesirable for mobile sensing in dynamic environments. This can be partially mitigated through hybridizing classical and quantum sensors, whereby the high-bandwidth classical sensors provide auxiliary information for (1) real-time feedback on the interferometer protocol to retain interferometer contrast and (2) post-correction of the measurement to ensure precision and accuracy \cite{Templier:2022,cheiney_navigation-compatible_2018,de_castanet_atom_2024}.

Open atom interferometry is an alternative to closed interferometry where the atomic matterwaves in the two interferometer arms are mismatched in position and momentum space at the final beamsplitter -- either deliberately or through platform dynamics -- resulting in spatially-dependent interference patterns (`spatial fringes') in the atomic density that encode the interferometer phase~\cite{Sugarbaker2013, Dickerson2013, Wigley2019}. The interferometer phase is extracted from a single image of the atomic density, enabling higher bandwidth acquisition and robustness to changes in the fringe bias and contrast -- both significant advantages for mobile operation.

However, open atom interferometers suffer two main limitations: first, a stable phase reference for the spatial fringe pattern is required; and second, the interferometer phase depends on the initial atomic velocity, which cannot be perfectly controlled in a dynamic setting. Thus, open atom interferometry has been limited to highly-controlled laboratories~\cite{Sugarbaker2013,Dickerson2013,Wigley2019,Asenbaum2020,overstreet2022,Wang2023} or to differential measurements where a phase reference is unnecessary and initial-velocity dependence can be cancelled exactly~\cite{Asenbaum2017}. A notable exception is Ref.~\cite{Mutinga2013}, which reported an open atom interferometry demonstration in microgravity; however, only contrast measurements were reported due to the lack of a stable phase reference~\footnote{Ref.~\cite{Mutinga2013} speculated that sensitivities of \SI{5e-12}{\meter/\second^2} could be reached with a suitable phase reference.}. Reference~\cite{Wigley2019} showed that an arbitrary pixel on the imaging camera can be used as a phase reference, which allowed an open interferometer to measure a gravitationally-induced phase. However, such a lab-frame reference was unstable on long timescales due to drifts in the camera location relative to the atomic source, making this approach unsuitable for mobile operation on dynamic platforms.

In this Letter, we report on the first demonstration of open interferometry with long-term phase stability for measuring gravity. Our approach employs a dual interferometric scheme \cite{yankelev_atom_2020,avinadav_composite-fringe_2020,bonnin_new_2018} where two open atom interferometers with interrogation times differing by $70\times$ are generated from the same atomic source and measured simultaneously. The resulting differential phase is independent of initial velocity contributions, allowing the interferometer to retain sensitivity to extrinsic fields and eliminating the requirement for an arbitrary (external) phase reference.  We demonstrate the efficacy of our method by monitoring gravitational tides over a period of $30$ hours using a Bose-condensed atomic source over a drop distance of \SI{23}{\centi\meter}, and achieving a precision of \SI{4.5}{\micro\gal} in approximately $2000$ shots. Our demonstration solves key challenges that have limited the widespread adoption of open interferometry, paving the way for future open interferometry experiments with mobile devices on dynamic platforms.

\par\emph{Open interferometry.---} Imperfect mode overlap at the output beamsplitter results in an open atom interferometer. In our work, this is achieved through temporal asymmetry $\delta T$ in the Mach-Zehnder (MZ) pulse sequence, such that $T$ is the time between the first beamsplitter and the mirror, and $T+\delta T$ is the time between the mirror and the second beamsplitter. For ideal $2n$-photon Bragg pulses~\cite{Muller:2008, Szigeti_2012, Altin2013} and average momentum $\bar{n}\hbar k$, this asymmetry gives rise to a phase shift~\cite{supp}:
\begin{align}\label{eq:phase_open_int}
   \phi =& (2nkg - \alpha)(T^2+\frac{1}{2}\delta T^2+2T\delta T+T_{\text{exp}}\,\delta T)\, \nonumber \\
    & + n(4\bar{n}\omega_r + 2kv_0 - \delta_0)\delta T,  
\end{align}
where $k$ is the wavenumber of the light pulse, $\omega_r = \hbar k^2 / (2m)$ is the atomic recoil frequency, $g$ is the acceleration due to gravity, $v_0$ is the initial atomic velocity at trap release, and $\texp$ is the duration between the release of the cloud from the trap and the first pulse.  To remain on resonance, the two photon detuning is set to $\delta(t) = \delta_0 + \alpha t$ assuming a value of $\delta_0$ at trap release and a frequency chirp rate $\alpha$. The key observation is that the contribution to the phase from the initial velocity and from gravity accrue at different rates, allowing them to be separated with an appropriate measurement protocol.

The phase of the interferometer is encoded in the spatial structure of the matterwave outputs, allowing readout while the outputs are still overlapped. This is a critical feature for mitigating readout delays in Bragg-based interferometry~\cite{Wigley2019}. However, using a camera pixel as the reference for phase extraction compromises both phase accuracy and stability, as does the initial velocity dependence of open interferometry. Shot-to-shot fluctuations in the cloud's average initial velocity and reference-frame position reduce precision, while long-term drifts in the reference-frame location lead to biases that compromise phase stability~\cite{Gaaloul2022}. Precise pre-release control of the atoms combined with velocity selection \cite{Cheng2018,Saywell2023} can substantially reduce the effect of initial velocity on the interferometer phase, but can add technological complexity and be difficult to implement on dynamic platforms. It has no effect on changes in the reference-frame location.

\begin{figure}[t]
    \centering
    \includegraphics[width=\linewidth]{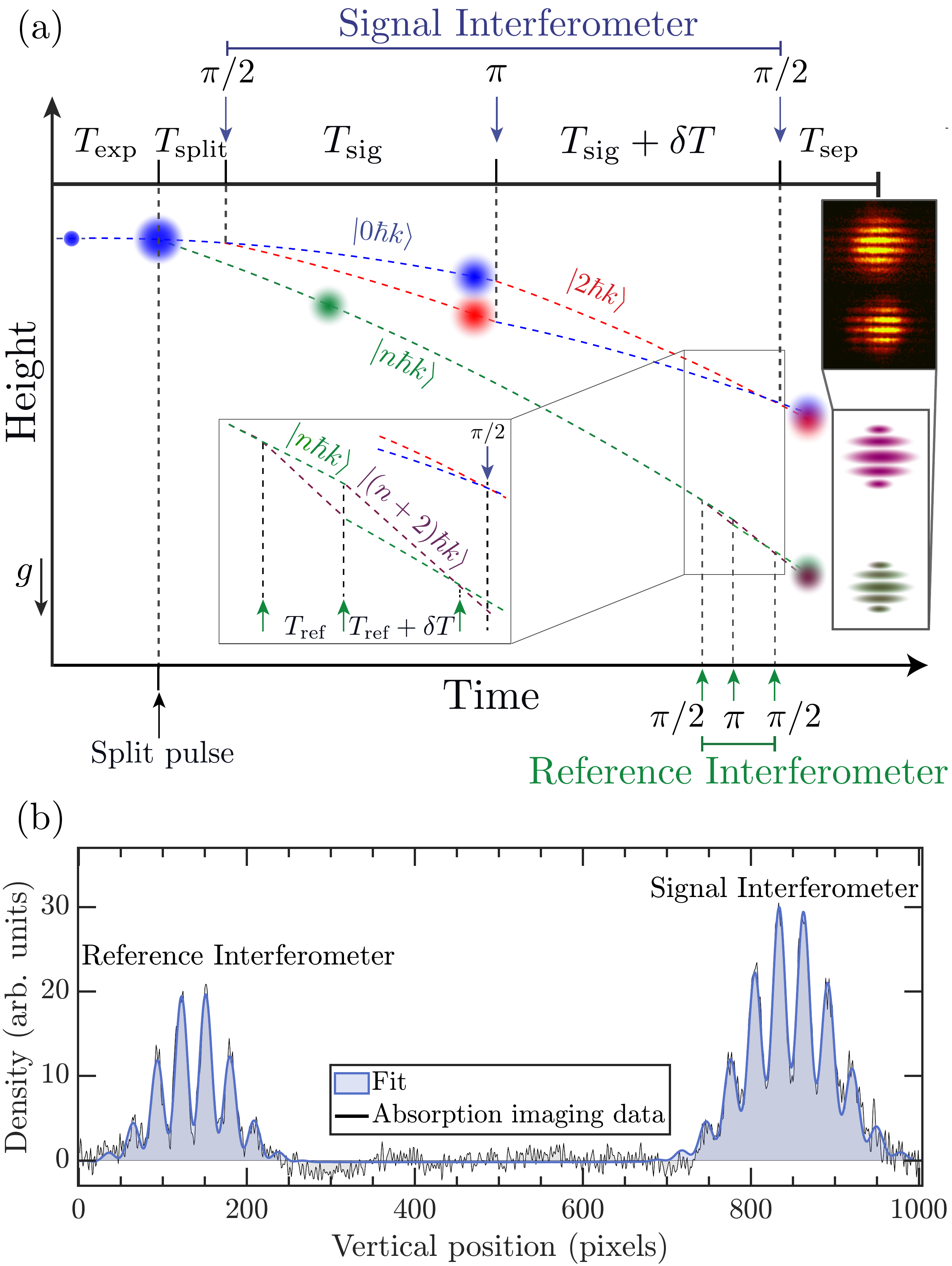}
    \caption{\small (a) Space-time diagram of dual open interferometry. The atom cloud is released from the trap and after expansion time $\texp$ a splitting pulse prepares two atom clouds in distinct momentum states which form their own independent interferometers. Following a brief period $\tsplit$, the signal interferometer sequence begins, which spans total duration $2\tsig + \delta T$. Concurrently, the reference interferometer undergoes a longer expansion time followed by a relatively short interferometry sequence. The spatial density distributions of the signal and reference interferometers are measured simultaneously after short separation times which differ slightly for both interferometers. (b) Vertical 1D atomic density measurement of both the reference and signal interferometer, extracted by summing a single 2D absorption image (insert shown in top right of (a)) in the horizontal direction and showing the spatial fringe patterns at both interferometer outputs.}
    \label{fg:Dual interferometer scheme}
\end{figure}

\emph{Dual open interferometry.---} Our approach employs a Bragg-based dual open interferometer, as shown in Fig.~\ref{fg:Dual interferometer scheme}. In this configuration, a signal interferometer of duration $2T_\text{sig} + \delta T$ accrues phase $\phi_{\text{sig}}$ due to both the initial velocity and gravitational acceleration. This is paired with a reference interferometer characterized by a shorter total duration $2T_\text{ref} + \delta T$, primarily sensitive to the initial velocity phase and, to a lesser extent, gravitational acceleration. By subtracting the phase of the signal interferometer from that of the reference, $\Delta\Phi_{\text{dual}} = \phi_{\text{sig}} - \phi_{\text{ref}}$, we cancel out the initial velocity effect and address the issue of drifting reference frames. This enables operation with a phase reference that is both fixed and well-defined in the atoms' frame. Specifically~\cite{supp},
\begin{align}\label{eq:dual_int}
\Delta\Phi_{\text{dual}} &= (2nkg - \alpha)\left(\tsig^2 - \tref^2 + \delta T\Delta \tsep\right) + \varphi,
\end{align} 
where $\Delta \tsep = (\tseps - \tsepl)$ is the difference in separation times, i.e. the interval between the last interferometer pulse and the imaging pulse for each interferometer. An additional phase $\varphi$ may be present depending on the timing of the reference interferometer relative to the signal interferometer; if the reference interferometer occurs wholly before or after the signal interferometer then $\varphi = 0$~\cite{supp}. Similarly to closed MZ interferometers, $g$ can be related to the measured phase $\Delta\Phi_{\rm dual}$ through atomic standards of length and time.

\par\emph{Experiment.---} We implemented the scheme illustrated in Fig.~\ref{fg:Dual interferometer scheme} using the laboratory-based atom interferometer previously reported in Refs.~\cite{Wigley2019,hardman2016,hardmanthesis}. The experiment, spanning approximately \SI{23}{\centi\meter} (\SI{216}{\milli\second} drop), begins with the production of a $^{87}$Rb Bose-Einstein condensate (BEC) in a crossed optical-dipole trap with $\mathord{\sim}10^6$ atoms in the $|F=1,m_F = -1\rangle$ state. Upon release from the trap, the atoms undergo \SI{10}{\milli\second} of freefall before being transferred to a magnetically-insensitive state $|F=1, m_F = 0\rangle$ using a microwave pulse sequence. This is followed by $\SI{57.05}{\milli\second}$ of expansion time to reduce mean-field energy and relocate the interferometry sequence away from strong stray magnetic fields caused by the ion pump. A splitting pulse then divides the initial BEC into two atom clouds in distinct momentum states to form separate interferometers. This differentiation is key for ensuring that each interferometer can be addressed independently. In our setup, the reference interferometer cloud is transferred from $\ket{0\hbar k}$ to $\ket{-4\hbar k}$ and operates between $\ket{-4\hbar k}$ and $\ket{-6\hbar k}$. Meanwhile, the signal interferometer operates between $\ket{0\hbar k}$ and $\ket{2\hbar k}$. Both initial clouds are then subjected to asymmetric MZ interferometer configurations with $T_\text{split}=\SI{0.5}{\milli\second}$, $T_\text{sig} = 70$~ms, $T_\text{ref} = 1$~ms, and $\delta T = \SI{0.45}{\milli\second}$ (see Fig.~\ref{fg:Dual interferometer scheme}), implemented via Gaussian pulses of FWHM durations of \SI{40}{\micro\second} and a one-photon detuning of approximately $\SI{-5.5}{\giga\hertz}$ from the D2 transition with a peak Rabi frequency of $2\pi\times \SI{6}{\kilo\hertz}$. A short $\tref$ ensures that the reference interferometer phase has a large contribution due to the initial velocity phase relative to the gravitational phase. The reference interferometer contains $\mathord{\sim}40\%$ of the total atom number in order to maintain an initial velocity sensitivity equivalent to the signal interferometer. After the interferometer sequences complete, the spatial density distributions of the signal and reference interferometers are measured concurrently after separation times of 8 and \SI{8.05}{\milli\second}, respectively~\footnote{The separation times are set so that the sinusoidal modulation of the output ports from each interferometer interferes constructively, allowing for readout with minimal separation while the ports overlap~\cite{Wigley2019}.}. The total drop time of both interferometers is identical: $\tdrop = 2T + \delta T + T_\text{sep} + \tsplit + \texp$.

\par\emph{Phase extraction and referencing.---} The temporal asymmetry $\delta T$ results in a spatial fringe pattern in the atomic density at each interferometer output, with a common spatial frequency $k_f= 2k\delta T/\tdrop$ and phase offsets $\phi_\text{sig}$ and $\phi_\text{ref}$ \cite{Krutzik2014Matter}.  We extract the phases by simultaneously capturing an absorption image of both interferometers' outputs on a single frame using a CCD camera~\footnote{Grasshopper2 GS2-FW-14S5M - 1384x1036 with a \SI{6.45}{\micro\meter} pixel size.}, and then fitting the vertical density profile to $G(z)\sin(k_f z + \phi_{\rm sig,ref})$ \cite{supp}.  We choose $G(z)$ to be Gaussian, although the exact form is unimportant provided it gives a good description of the vertical density.

Imaging both interferometers on a single frame provides the crucial common phase reference needed to maintain a constant relative phase, since any change in phase affecting one interferometer similarly impacts the other. Without this, phase referencing is highly susceptible to displacements in the camera pixel, and therefore to vibrations and temperature changes. In a stable laboratory, pixel referencing is mainly susceptible to slow temperature-induced changes over extended periods. However, on dynamic platforms where rapid and significant temperature changes and vibrations are present, these factors could introduce a significant bias. Even a \SI{1}{\micro\meter} change in the reference-frame position would give a \SI{45}{\micro\gal} error for our experiment's parameters, underscoring the effectiveness of our method.

\begin{figure}[t]
    \centering
        \includegraphics[width=\linewidth]{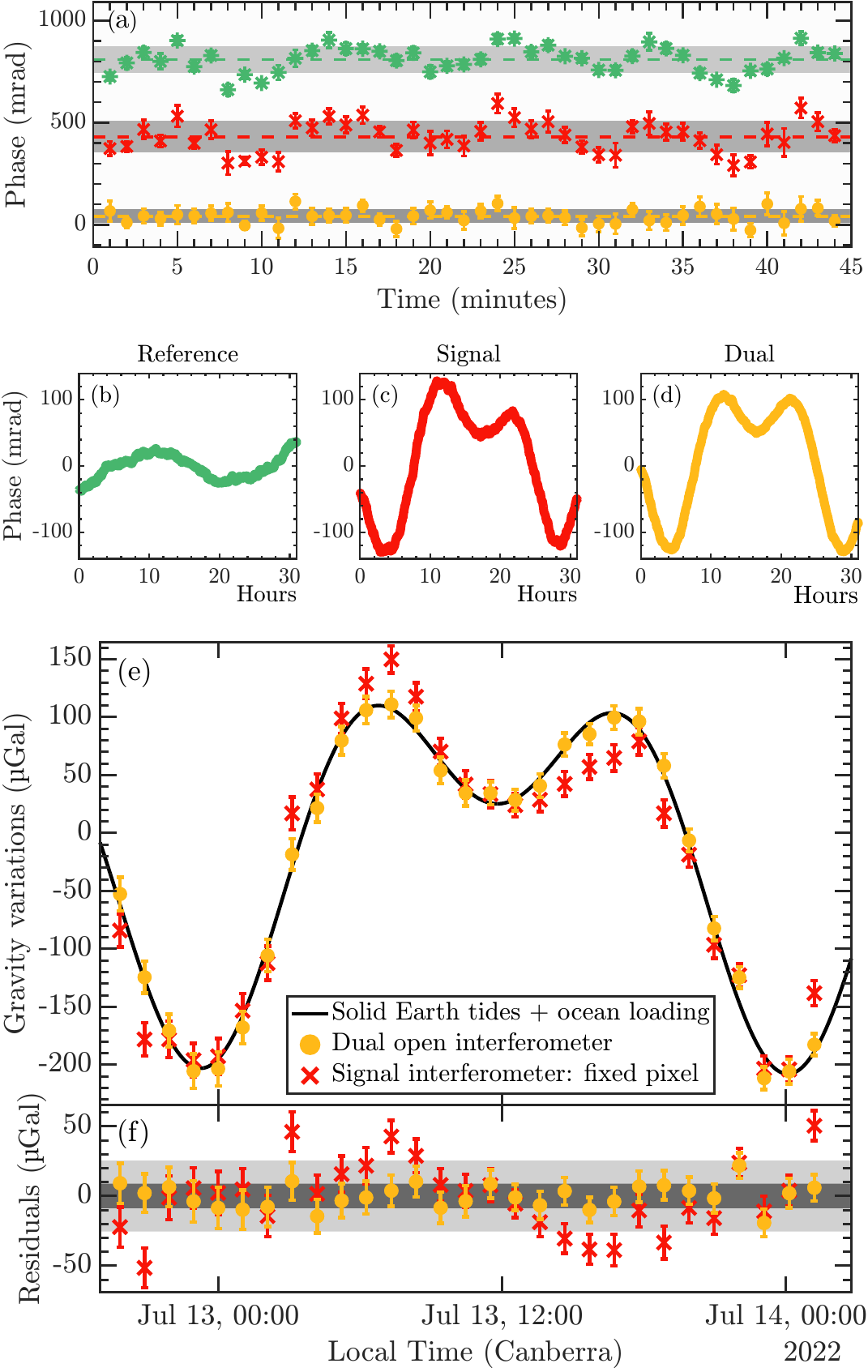}
    \caption{\small (a) Short-term phase stability measurements for the dual interferometer (yellow circles) compared to the individual signal (red crosses) and reference (green stars) interferometers. Data from each interferometer are offset for visual clarity. The dark grey shading indicates the standard deviation for each interferometer dataset. (b)-(d) 800-point moving average of reference, signal, and dual interferometer phase measurements obtained over $\sim\!\SI{30}{\hour}$ duration. (e) Comparison of the solid Earth tide (black line), dual interferometer (yellow circles) and signal interferometer (red crosses) with data averaged over 1 hour bins. (f) Residuals resulting from subtracting the solid Earth tide from the measurements of both interferometers. The standard error of the residuals for the dual interferometer (dark shading) is $2.5 \times$ smaller than for the signal interferometer (light shading).}
    \label{fg:tides}
\end{figure}
\begin{figure}[t]
\centering
\includegraphics[width=\columnwidth]{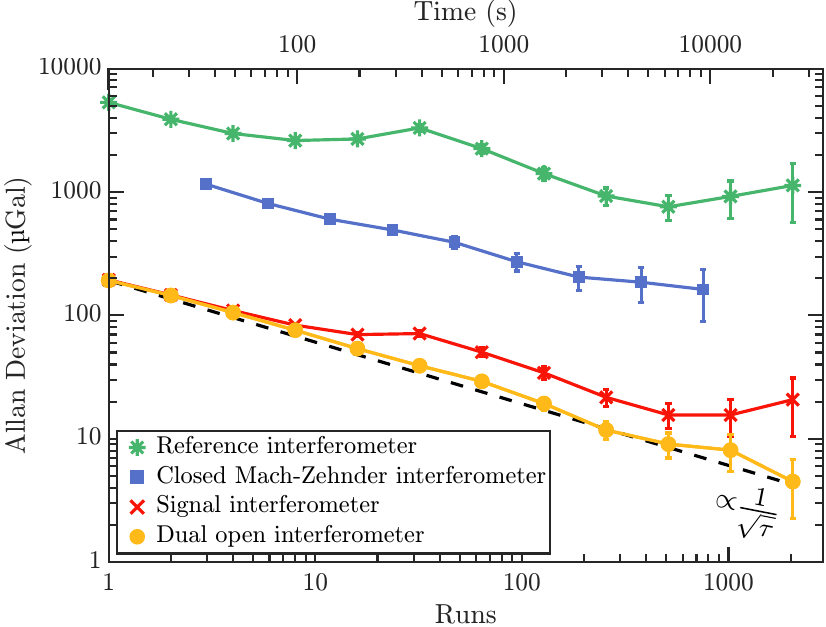}
{\caption{\small Allan deviations of the gravitational acceleration signal corrected for the solid Earth tide are shown for the dual interferometer (yellow circles), signal interferometer (red crosses), reference interferometer (green stars), and a closed MZ interferometer (blue squares). The closed MZ operates at shorter interrogation time $T = \SI{15}{\milli\second}$, necessary to obtain resolvable output ports within the $\SI{23}{\centi\meter}$ height constraints of our experiment. The dashed black line plots $\tau^{-1/2}$ scaled by the single-shot sensitivity of the dual interferometer. Each run is approximately \SI{12}{\second}.{\label{fg:adev}}}}
\end{figure}  
\par\emph{Short-term phase stability.---} To demonstrate short-term stability, we performed a series of phase measurements as shown in Fig.~\ref{fg:tides} (a). We extracted the phases of the reference interferometer and the signal interferometer over $\mathord{\sim}\SI{45}{\min}$ (220 shots). Subtracting the signal phase from the reference interferometer phase, we calculated the dual open interferometer phase. While both the reference and signal phases exhibit significant run-to-run fluctuations, the fluctuations are correlated ($0.88$ correlation coefficient) since they are partially caused by changes in the initial atomic velocity and are significantly reduced on subtraction. This method demonstrates a twofold improvement in phase stability compared to the signal interferometer alone.

\par\emph{Long-term phase stability.---} Fluctuations in initial velocity and phase reference over long periods, which do not average to zero, introduce bias and degrade the long-term stability and accuracy of the interferometer. To demonstrate our method's long-term phase stability and its ability to counteract these biases, we tracked tidal variations in local gravity over a 30-hour period, taking measurements every 12 seconds. The procedure for measuring phase is detailed in Figs.~\ref{fg:tides} (b-c-d). Figures~\ref{fg:tides} (b) and (c) display phase data for the reference and signal interferometers, respectively, each smoothed with an 800-point moving average. Fig.~\ref{fg:tides} (d) illustrates the dual interferometer's phase reconstruction, achieved by subtracting the reference phase from the signal phase and then applying the same moving average.

Our gravity measurements, taken during a king tide, were anticipated to yield a symmetrical tidal phase profile. Yet, the signal interferometer erroneously exhibited strong asymmetry, indicative of a neap tide. In contrast, the dual interferometer consistently revealed a symmetrical pattern, aligning with the expected profile of a king tide. This qualitative analysis clearly demonstrates the impact of long-term drifts, and showcases the dual open interferometer's ability to correct for these long-term drifts observed in the signal interferometer.

In Figs.~\ref{fg:tides} (e-f), we compare our measurements to a solid Earth tides model and ocean loading \cite{tidalmodel} with each data point representing a one hour average. As seen in Fig.~\ref{fg:tides} (e), the signal interferometer, which uses a fixed pixel as a reference, shows some alignment with the tidal model but also reveals observable drifts. Conversely, the dual interferometer exhibits excellent agreement with the tidal model, with a standard error of the residuals of $\SI{8.6}{\micro\gal}$. The residuals in Fig.~\ref{fg:tides} (f) of the dual open interferometer against the tidal model showcase over $2.5\times$ improvement compared to the signal interferometer. 

To further assess the overall phase stability of our scheme, we performed an Allan deviation analysis \cite{adev} on the different interferometer measurements. As shown in Fig.~\ref{fg:adev}, both the signal and dual interferometers exhibit similar single-shot sensitivities of approximately \SI{200}{\micro\gal}, with the dual interferometer slightly better by less than \SI{10}{\micro\gal}. For short integration times (around 10--20 runs), both interferometers' sensitivities follow a $\tau^{-1/2}$ trend. This is consistent with Ref.~\cite{Wigley2019}, which reported comparable phase stability using a pixel reference over tens of runs, each with a \SI{12}{\second} duty cycle. Beyond this timescale, the signal interferometer begins to drift, whereas the dual interferometer keeps integrating along the $\tau^{-1/2}$ trend, to reach a sensitivity of \SI{4.5}{\micro\gal} in approximately $2000$ runs, compared to \SI{21}{\micro\gal} for the signal interferometer. Given there is no deviation in the $\tau^{-1/2}$ scaling of the dual interferometer, it is possible that long-term stability is maintained beyond the $\mathord{\sim}\SI{30}{\hour}$ data collection period, which if true would have allowed a precision exceeding \SI{4.5}{\micro\gal}.

\emph{Comparison to closed MZ.---}In our open MZ protocol, imaging is performed shortly after the final beamsplitter while the interferometer modes are spatially overlapping. This provides a significant advantage over closed MZ Bragg-based interferometers, which require the momentum modes to be spatially separated at the detector~\cite{Wigley2019}. Consequently, for fixed drop length, an open MZ protocol can operate with a longer interrogation time -- and thus better sensitivity -- compared to a closed MZ interferometer. In our \SI{23}{\centi\meter} drop, the requirement for the interferometer modes to spatially separate after the final beamsplitter limits the interrogation time of a closed MZ to $T=\SI{15}{\milli\second}$. Figure~\ref{fg:adev} demonstrates a sixfold improvement in single-shot sensitivity between the closed and open MZ schemes, due entirely to the $T^2$ dependence of the scale factor~\cite{Kritsotakis2018}. Although one would expect an increase in sensitivity of $(70/15)^2 \approx 22$, as $T$ increases the interferometer sensitivity becomes limited by the motion of our retroreflecting mirror.  If we were to operate both protocols at $T = \SI{70}{\milli\second}$, they would have similar sensitivities, but the closed MZ interferometer would require a drop length of $\mathord{\sim}\SI{1}{\meter}$ rather than \SI{23}{\centi\meter}.

Other strategies exist to mitigate the readout delay in Bragg-based interferometers, such as velocity-selective Bloch separator pulses~\cite{Piccon2022} and Raman labelling \cite{Cheng2018}~\footnote{Direct measurements of the momentum distribution could also mitigate readout delay, although detection requirements are very stringent~\cite{Kritsotakis2018}.}. These can significantly improve sensitivity for a given device size at the cost of additional experimental complexity and increased susceptibility to phase errors induced by noisy environments. Due to the cloud momentum width, these pulses are susceptible to phase errors arising from cross-coupling between spatially-overlapping momentum modes and also suffer from the potential for different velocity classes to carry different phases~\cite{Piccon2022}. In contrast, our open interferometer avoids the need to isolate the interferometer outputs in momentum space entirely by directly extracting the phase from the atomic cloud's spatial interference pattern.

\par\emph{Conclusion and outlook.---} The dual open interferometer scheme we have introduced represents a significant step forward in the field of open atom interferometry, marking the first demonstration of long-term stability in an open configuration. Our method enabled continuous monitoring of local gravity and its tidal variations over a nearly 30-hour period. Conducted within a height of \SI{23}{\centi\meter}, this allowed interferometer times about 4.5 times longer than those achievable with a closed Bragg MZ interferometer, leading to a sixfold improvement in single-shot sensitivity for the same device size. The ability of our approach to consistently extract phase data in every run makes it particularly useful for dynamic and mobile applications, where additional measurements needed to measure interferometer contrast are often impractical. This attribute is equally beneficial for long-drop experiments constrained by limited operational runs, circumventing the need for extensive phase reconstruction processes. Although our demonstration used a Bose-condensed source, this was a choice of convenience rather than necessity; we expect dual open interferometry to be performant for cold (i.e. non-condensed) thermal and laser-cooled sources. Indeed, dual open interferometry should provide robustness to velocity-spread effects that degrade contrast in closed MZ interferometers.

The use of a reference interferometer in our setup effectively addresses issues of unreliable phase referencing due to initial velocity fluctuations and environmental changes. This development opens avenues for a wide range of applications in open interferometry. For instance, it could enable shot-noise limited operation in a wider range of cold-atom sensors, expanding the use-cases of existing devices~\cite{Geiger:2020,Narducci2022}. Our open interferometry protocol could be combined with classical inertial sensors to provide additional robustness in dynamic environments, similar to hybrid sensing with closed MZ protocols~\cite{Templier:2022,de_castanet_atom_2024}. Immediate future work should aim to validate our approach in mobile cold-atom devices on dynamic platforms, and extend our scheme to the measurement of other quantities such as rotations, gravity gradients and magnetic field gradients and curvature. An intriguing possibility is the simultaneous measurement of several quantities in a single experimental run, e.g. by simultaneously running dual open interferometers in multiple magnetic substates~\cite{hardman2016}. Thus, our work not only marks a significant step in open atom interferometry but also lays the groundwork for the development of versatile, all-in-one quantum sensors~\cite{Barrett:2019}.

\par\emph{Acknowledgements.---}
We acknowledge the Ngunnawal people as the traditional custodians of the land on which this research was conducted. We acknowledge their continuing connection to the land, waters, and community, and acknowledge that sovereignty was never ceded. We thank Cass Sackett, Kaiwen Zhu, Raphael Piccon and Kevin Stitley for insightful discussions. 
This research was funded by the Australian Research Council project DP190101709. SAH acknowledges support through an Australian Research Council Future Fellowship Grant No. FT210100809. SSS was supported by an Australian Research Council Discovery Early Career Researcher Award (DECRA), Project No. DE200100495. 
\bibliography{references}

\clearpage

\newcommand\recoil{\omega_{\rm r}}
\renewcommand{\theequation}{S.\arabic{equation}}
\setcounter{section}{0} 

\newcommand{\tsigpulselast}{T_{\text{3,sig}}}
\newcommand{\tsigpipulse}{T_{\text{2,sig}}}
\newcommand{\trefpulselast}{T_{\text{3,ref}}}
\newcommand{\trefpipulse}{T_{\text{2,ref}}}
\newcommand{\detref}{\delta_{\text{ref},0}}
\newcommand{\detsig}{\delta_{\text{sig},0}}
\newcommand{\phisig}{\phi_{\text{sig}}}
\newcommand{\phiref}{\phi_{\text{ref}}}
\newcommand{\barnref}{\bar{n}_{\text{ref}}}
\newcommand{\barnsig}{\bar{n}_{\text{sig}}}
\newcommand{\kfringe}{k_{\rm fringe}}
\renewcommand{\i}{\mathrm{i}}
\widetext
\begin{center}
\textbf{\large Supplementary Material: A Dual Open Atom Interferometer for Compact, Mobile Quantum Sensing}
\end{center}
In this supplemental material we provide (1) a derivation of the phase for an open Mach-Zehnder Bragg-pulse atom interferometer, (2) a derivation of the phase for our dual open interferometer scheme, and (3) the explicit form of the fit used to extract the phase from our spatial fringe absorption images.

\section{Derivation of open Mach-Zehnder interferometer phase shift, Eq.~(1)}
The interferometer described in the main text uses two-photon Bragg transitions to coherently couple different momentum states of individual atoms while leaving the internal state unchanged. For counter-propagating lasers with a mean wavenumber $k$ and frequency difference $\delta$, the temporal evolution of the position-space wavefunction $\psi(x,t)$ in one dimension is [68,69]
\begin{equation}
    \i\hbar \frac{\partial\psi(x,t)}{\partial t} = -\frac{\hbar^2}{2m}\frac{\partial^2\psi(x,t)}{\partial x^2} + \hbar\left(\Omega e^{2\i k x - \i\delta t} + \Omega^* e^{-2\i k x +\i\delta t}\right)
\end{equation}
for two-photon Rabi frequency $\Omega$. Note that we have neglected the AC Stark shift as it is common to all momentum states. Changing variables ($x\to x - \delta t/(2k)$), and transforming to momentum space, we have
\begin{equation}
    \i\hbar \frac{\partial\psi(p,t)}{\partial t} = \left[\frac{p^2}{2m} - \frac{\delta}{2k}p\right]\psi(p,t) + \hbar\Omega\psi(p + 2\hbar k,t) + \hbar\Omega^*\psi(p - 2\hbar k,t),
\end{equation}
which shows that the counter-propagating laser beams couple momentum states that differ by $2\hbar k$. Defining $c_n(p,t) \equiv \psi(p + 2n\hbar k,t)$, we obtain a system of coupled equations
\begin{equation}
    \i\dot{c}_n(p,t) = \left[\hbar^{-1}\left(\frac{p^2}{2m} - \frac{\delta}{2k}p\right) + 4n^2\recoil + n\left(\frac{2kp}{m} - \delta\right)\right]c_n(p,t) + \Omega c_{n+1}(p,t) + \Omega^*c_{n-1}(p,t),
\end{equation}
where $\recoil = \hbar k^2/(2m)$ is the recoil frequency. The effective detuning for transitions between initial momentum state $\ket{p + 2 n_i \hbar k}$ and final momentum state $\ket{p + 2 n_f \hbar k}$ is then
\begin{equation}
    \xi = n\left(4\bar{n}\recoil + 2kv - \delta\right)
\end{equation}
with $v = p/m$, and where we have defined $n = n_f - n_i$ as the momentum transfer order and $\bar{n} = n_f + n_i$ such that $\bar{n}\hbar k$ is the average imparted momentum. For gravimetry, $\delta$ is typically parametrized as $\delta = \delta_0 + \alpha t$, where $\delta_0$ is the initial detuning and $\alpha$ is a frequency sweep rate. For atoms falling under gravitational acceleration $g$, $v(t) = v_0 + g t$, which means that in order to remain close to resonance we require $\alpha = 2k\hat{g}$, where $\hat{g}$ is an estimate of $g$. The time-dependent atom-light detuning is then
\begin{equation}
    \xi(t) = 2nk\delta g\, t + n\left(4\bar{n}\recoil + 2kv_0 - \delta_0\right),
\end{equation}
where $\delta g = g - \hat{g}$ is the error in the estimate of gravity.

We now assume we have an open Mach-Zehnder interferometer with a $\pi/2$-$\pi$-$\pi/2$ pulse sequence where pulses occur at times $T_1 = \texp$, $T_2 = \texp + T$, and $T_3 = \texp + 2T + \delta T$, where $T$ is the interrogation time and $\delta T$ is the temporal asymmetry. Neglecting phase evolution during the pulses, the phase shift after the pulse sequence is
\begin{align}
	\phi &= \int_{T_2}^{T_3} \xi(t) dt - \int_{T_1}^{T_2} \xi(t) dt\notag\\
	&= 2nk\delta g \left(T^2 + 2T\delta T + \frac{1}{2}\delta T^2 + \texp\delta T\right) + n\left(4\bar{n}\recoil + 2kv_0 - \delta_0\right)\delta T,
	\label{eq:open-phase-shift}
\end{align}
where the gravitationally-induced phase scales with the momentum transfer $2n \hbar k$, and the temporal asymmetry introduces phase sensitivity to the velocity of the atoms at the first pulse, $v(\texp) = v_0 + g\texp$. The initial two-photon detuning is chosen to eliminate the second term in Eq.~\eqref{eq:open-phase-shift}, i.e. $\delta_0 = 4\bar{n}\recoil + 2k\hat{v}_0$, where $\hat{v}_0$ is an estimate of the initial velocity, and is typically but not necessarily $\hat{v}_0 = 0$. The sensitivity of the interferometer phase to initial velocity has two effects. First, ensembles with a spread in initial velocities will acquire velocity-dependent phase shifts that map to a position-dependent phase shift after ballistic expansion, giving rise to spatial fringes which can be used for enhanced readout as detailed in the main text. Second, the interferometer phase can change in response to changes in the \textit{mean} velocity of the sample that are uncompensated by corresponding changes in the estimate $\hat{v}_0$, which is typically a fixed value.

\section{Derivation of dual open interferometer phase shift, Eq.~(2)}
The dual open interferometry method eliminates the sensitivity to changes in the mean initial velocity by generating two interferometers from the same initial source that have different interferometer times. Suppose that we have one interferometer, denoted the signal intererometer, with interferometer time $\tsig$ and average momentum order $\barnsig$, and a second interferometer, denoted the reference interferometer, with interferometer time $\tref$ and average momentum order $\barnref$. The momentum transfer order $n$ is assumed to be the same. We start by assuming that the reference interferometer occurs entirely after the signal interferometer is finished; i.e., the first $\pi/2$ pulse of the reference interferometer occurs after the last $\pi/2$ pulse of the signal interferometer. We further assume that the outputs of both interferometers are measured at the same total drop time $T_{\rm drop}$ which implies the following relationship
\begin{equation}
	T_{\rm drop} = \texpL + 2\tsig + \delta T + \tsepl = \texps + 2\tref + \delta T + \tseps,
\end{equation}
for initial $\pi/2$ pulse times of $\texpL$ (signal) and $\texps$ (reference), and similarly for the separation times $\tsepl$ and $\tseps$.  We then have the following equality:
\begin{equation}
	\tsepl - \tseps = -2(\tsig - \tref) - (\texpL - \texps).
\end{equation}
In order to resonantly address each interferometer, the detuning offsets $\delta_{(\textrm{sig},\textrm{ref}),0}$ are
\begin{equation}
    \delta_{(\textrm{sig},\textrm{ref}),0} = 4\bar{n}_{(\textrm{sig},\textrm{ref})}\recoil + 2k\hat{v}_0
\end{equation}
which share a common estimate of the mean velocity $\hat{v}_0$. The phase difference $\Delta \Phi_{\rm dual} = \phisig - \phiref$ is then
\begin{align}
	\Delta \Phi_{\rm dual} &= 2nk\delta g\left[\tsig^2 - \tref^2 + \left(2[\tsig - \tref] + [\texpL - \texps]\right)\delta T + \frac{1}{2}\delta T^2\right]  \\ &+ n\left[(4\barnsig\recoil - \detsig) - (4\barnref\recoil - \detref)\right]\delta T\notag\\
	&= 2nk\delta g\left[\tsig^2 - \tref^2 - (\tsepl - \tseps)\delta T\right].\label{eq:dual-interferometer-phase}
\end{align}
Since the two interferometers are generated from the same source with the same initial velocity, the phase shift associated with the initial velocity is cancelled by the subtraction of the two phases.

We now consider the situation where the reference interferometer occurs entirely prior to the end of the signal interferometer, such that $\tsigpipulse < \texps,\trefpulselast < \tsigpulselast$.  In this case, the detuning of the laser must be changed during the signal interferometer so that it resonantly addresses the reference interferometer.  The atom-laser detuning is then
\begin{align}
	\xi(t) &= n\left[4\barnsig\recoil + 2kv_0 + 2kgt - \delta(t)\right]\notag\\
	&= n\left[4\barnsig\recoil + 2kv_0\right] + nk\delta g t - n\begin{cases}
		\detsig, & \texpL < t < \texps\\
		\detref, & \texps < t < \trefpulselast\\
		\detsig, & \trefpulselast < t < \tsigpulselast
	\end{cases}\\
	&= \begin{cases}
			\xi_s(t), & \texpL < t < \texps\\
			\xi_r(t), & \texps < t < \trefpulselast\\
			\xi_s(t), & \trefpulselast < t < \tsigpulselast.
		\end{cases}
\end{align}
The phase of the reference interferometer remains unchanged, but the phase of the signal interferometer is different from the previous situation where the reference interferometer occurs after the signal interferometer.  The phase of the signal interferometer is now
\begin{align}
	\phisig &= \int_{\tsigpipulse}^{\tsigpulselast} \xi(t) dt - \int_{\texpL}^{\tsigpipulse} \xi(t) dt\notag\\
	&= \int_{\tsigpipulse}^{\tsigpulselast} \xi_s(t) dt - \int_{\texpL}^{\tsigpipulse} \xi_s(t) dt + \int_{\texps}^{\trefpulselast} \xi_r(t) - \xi_s(t) dt
\end{align}
where the last term is
\begin{align}
	\int_{\texps}^{\trefpulselast} \xi_r(t) - \xi_s(t) dt &= n\left[\detsig - \detref\right](2\tref + \delta T)\notag\\
	&= 4n\recoil\left[\barnsig - \barnref\right](2\tref + \delta T).
\end{align}
The dual interferometer phase is then
\begin{align}
	\Delta \Phi_{\rm dual} &= 2nk\delta g\left[\tsig^2 - \tref^2 - (\tsepl - \tseps)\delta T\right] + 4n\recoil\left[\barnsig - \barnref\right](2\tref + \delta T)
	\label{eq:dual-interferometer-phase-interior}
\end{align}
where the second term in Eq.~\eqref{eq:dual-interferometer-phase-interior} is due to the frequency change during the signal interferometer.
\section{Fit used to extract phase from the spatial fringe pattern}
As described in the main text, in each run the two interferometers are imaged simultaneously onto a CCD camera using absorption imaging, with both interferometer outputs captured within a single frame. This provides a 2D column density; after integrating over the dimension transverse to the vertically-oriented interferometry beams, we obtain a 1D atomic density distribution along the vertical direction. To extract the phases $\phi_\text{sig}$ and $\phi_\text{ref}$, we fit the following function to this distribution:
\begin{align}\label{eq:fit}
f(z) &= A_{\text{ref}}\exp\left[-\frac{(z - z_{\text{ref}})^2}{2\sigma_{\text{ref}}^2}\right]\left[1-B_{\text{ref}}\sin(\kfringe\,(z - z_0)-\phi_{\text{ref}})\right]
\\ &+C +A_{\text{sig}}\exp\left[-\frac{(z - z_{\text{sig}})^2}{2\sigma_{\text{sig}}^2}\right]\left[1-B_{\text{sig}}\sin(\kfringe\,(z - z_0)-\phi_{\text{sig}})\right],
\end{align}
where $ A_{\text{ref}}, A_{\text{sig}}, z_{\text{ref}}, z_{\text{sig}}, \sigma_{\text{ref}}, \sigma_{\text{sig}}, B_{\text{ref}}, B_{\text{sig}}, C,\phi_{\text{ref}}$ and $\phi_{\text{sig}}$ 
are free parameters. The spatial frequency of the fringes, $\kfringe = nk\delta T/\tdrop$, is known $\emph{a priori}$, and in fact is used to calibrate the magnification of the imaging system through multiple runs; afterwards, it remains fixed. The phase reference $z_0$ can be set arbitrarily provide it is identical for both clouds. Typically, we set $z_0$ to the position at beginning of the measurement record. 


{\small
\begin{enumerate}
	\setlength\itemsep{0.05em}
	\setcounter{enumi}{67}
	\item P. B. Blakie and R. J. Ballagh, ``Mean-field treatment of Bragg scattering from a Bose-Einstein condensate'', Journal of Physics B: Atomic, Molecular and Optical Physics \textbf{33}, 3961 (2000).
 \setcounter{enumi}{68}
	\item S. S. Szigeti, ``Controlled Bose-Condensed Sources for Atom Interferometry'', Ph.D. thesis, The Australian National University (2013).
\end{enumerate}
}

\end{document}